\documentclass[12pt,a4paper]{article}
\def\bild#1#2{    
        \vspace*{-5mm}
        \begin{center}
          \includegraphics[width=#2cm]{#1}
        \end{center}
        }
\usepackage{axodraw}
\usepackage{graphicx}
\newcommand{\vs}{\vspace{-0.25cm}}

\begin{document}

\begin{center}

{\Large
\textbf{Nuclear energy density functional\\ from chiral pion-nucleon 
dynamics}\footnote{Work supported in part by BMBF, GSI and DFG.} }

\bigskip
N. Kaiser$\,^a$, S. Fritsch$\,^a$ and W. Weise$\,^{a,b}$\\

\bigskip

{\small $^a$\,Physik Department, Technische Universit\"{a}t M\"{u}nchen, 
D-85747 Garching, Germany\\

\smallskip

$^b$\, ECT$^*$, I-38050 Villazzano (Trento), Italy\\

\smallskip

{\it email: nkaiser@physik.tu-muenchen.de}}

\end{center}

\bigskip

\begin{abstract}
We calculate the nuclear energy density functional relevant for N=Z even-even 
nuclei in the systematic framework of chiral perturbation theory. The 
calculation includes the one-pion exchange Fock diagram and the iterated
one-pion exchange Hartree and Fock diagrams. From these few leading order
contributions in the small momentum expansion one obtains already a very good 
equation of state of isospin symmetric nuclear matter. We find that in the
region below nuclear matter saturation density the effective nucleon mass 
$\widetilde M^*(\rho)$ deviates by at most $15\%$ from its free space value 
$M$, with $0.89M<\widetilde M^*(\rho)<M$ for $\rho < 0.11 \,{\rm fm}^{-3}$ and 
$\widetilde M^*(\rho)>M$ for higher densities. The parameterfree strength of 
the $(\vec\nabla \rho)^2$-term, $F_\nabla(k_f)$, is at saturation density 
comparable to that of phenomenological Skyrme forces. The magnitude of 
$F_J(k_f)$ accompanying the squared spin-orbit density $\vec J\,^2$ comes out
somewhat larger. The strength of the nuclear spin-orbit interaction, $F_{so}(
k_f)$, as given by iterated one-pion exchange is about half as large as the
corresponding empirical value, however, with the wrong negative sign. The
novel density dependencies of $\widetilde M^*(\rho)$ and $F_{\nabla,so,J}(k_f)$
as predicted by our parameterfree calculation should be examined in nuclear 
structure calculations (after introducing an additional short range spin-orbit
contribution constant in density).  
\end{abstract}

\bigskip


\bigskip

\bigskip 

\section{Introduction}
Among the various phenomenological interactions that have been used extensively
in the description of nuclei, the Skyrme force \cite{skyrme} has gained much
popularity because of its analytical simplicity and its ability to reproduce 
nuclear properties over the whole periodic table within the self-consistent 
Hartree-Fock approximation. Several Skyrme parameterizations have been tailored
to account for single particle spectra \cite{sk3}, giant monopole resonances 
\cite{skm} or fission barriers of heavy nuclei \cite{skmstar}. Recently, a new 
Skyrme force which also reproduces the equation of state of pure neutron matter
up to neutron star densities, $\rho_n \simeq 1.5 \,{\rm fm}^{-3}$, has been 
proposed in ref.\cite{sly} for the study of nuclei far from stability. A 
microscopic interpretation of the various parameters entering the effective 
Skyrme forces is generally put aside. Sometimes the energy density functional 
is just parameterized without reference to any effective (zero-range) two-body
interaction in order to avoid the complete fixing of time-reversal-odd terms by
the Pauli-exclusion principle \cite{reinhard}.   

Another widely and successfully used approach to nuclear structure calculations
are relativistic mean-field models \cite{ringreview}. In these 
models the nucleus is described as a collection of independent Dirac-particles 
moving in self-consistently generated scalar and vector mean-fields. The 
footprints of relativity become visible through the large nuclear spin-orbit 
interaction which emerges in that framework naturally from the interplay of the
two strong and counteracting (scalar and vector) mean-fields. The corresponding
many-body calculations are usually carried out in the Hartree approximation,
ignoring the negative-energy Dirac-sea. For a recent review on self-consistent 
mean-field models for nuclear structure, see ref.\cite{reinhard}. In that 
article the relationship between the relativistic mean-field models and the 
Skyrme phenomenology is also discussed.  

The first conditions to be fulfilled by any phenomenological nucleon-nucleon 
interaction come from the (few empirically known) properties of infinite 
nuclear matter. These are the saturation density $\rho_0 =2k_{f0}^3/3\pi^2$, 
the binding energy per particle $-\bar E(k_{f0})$ and the compression modulus 
$K =k_{f0}^2 \bar E''(k_{f0})$ of isospin symmetric nuclear matter as well as 
the asymmetry energy $A(k_{f0})$. In general, Skyrme forces involve several 
more parameters related to terms in the energy density functional which vanish
identically in homogeneous nuclear matter (like the spin-orbit coupling
proportional to the density-gradient).

In a recent work \cite{nucmat}, we have used chiral perturbation theory for a  
systematic treatment of the nuclear matter many-body problem. In this 
calculation the contributions to the energy per particle, $\bar E(k_f)$,
originate exclusively from one- and two-pion exchange between nucleons and they
are ordered in powers of the Fermi momentum $k_f$ (modulo functions of
$k_f/m_\pi$). It has been demonstrated in ref.\cite{nucmat} that the empirical 
saturation point $(\rho_0 \simeq 0.17\,{\rm fm}^{-3}\,, \bar E(k_{f0})\simeq
-15.3\,{\rm MeV})$ and the nuclear matter compressibility $K\simeq 255\,$MeV 
can be well reproduced at order ${\cal O}(k_f^5)$ in the small momentum 
expansion with just one single momentum cut-off scale of $\Lambda \simeq 0.65
\,$GeV which parameterizes the necessary short range NN-dynamics. Most 
surprisingly, the prediction for the asymmetry energy, $A(k_{f0})=33.8\,$MeV, 
is in very good agreement with its empirical value. Furthermore, as a 
nontrivial fact, pure neutron matter is predicted to be unbound and the 
corresponding equation of state $\bar E_n(k_n)$ agrees roughly with that of 
sophisticated many-body calculations for low neutron densities $\rho_n \leq 
0.25\,$fm$^{-3}$. In a subsequent work \cite{pot}, the momentum and density 
dependent (real) single-particle potential $U(p,k_f)$ has been calculated in 
the same framework. It was found that chiral $1\pi$- and $2\pi$-exchange give 
rise to a potential depth for a nucleon at the bottom of the Fermi sea of 
$U(0,k_{f0})=-53.2\,$MeV. This value is in very good agreement with the depth 
of the empirical optical model potential and the nuclear shell model potential.
In fact very similar results for nuclear matter can be obtained already at 
order ${\cal O}(k_f^4)$ in the small momentum expansion (by dropping the 
relativistic $1/M^2$-correction to $1\pi$-exchange and the irreducible 
$2\pi$-exchange of order ${\cal O}(k_f^5)$) with a somewhat reduced cut-off
scale of $\Lambda \simeq 0.61\,$GeV. Detailed results of this ${\cal O}(k_f^4)
$-calculation can be found in ref.\cite{lutzpot} (it is named "mean-field 
treatment of the NN-contact interaction" therein).  

Given the fact that many properties of nuclear matter can be well described by
chiral $\pi N$-dynamics treated perturbatively up to three-loop order it is
natural to consider in a further step the energy density functional relevant
for inhomogeneous many-nucleon systems (i.e. finite nuclei). We will restrict
ourselves here to the isospin symmetric case of equal proton and neutron 
number N=Z. The aim of this work is to calculate density-dependent 
generalizations of those combinations of Skyrme parameters which belong to 
terms specific for the inhomogeneous many-nucleon system, such as
$(\vec \nabla \rho)^2$. This novel density-dependence is a consequence of the 
finite range character of the $1\pi$- and $2\pi$-exchange interaction. We 
stress already here that our results for the density-dependent strength
functions are completely parameterfree. In particular, they are independent of 
the cut-off scale $\Lambda$ adjusted in ref.\cite{nucmat,lutzpot} to the 
binding energy per particle $-\bar E(k_{f0})=15.3\,$MeV. 

Our paper is organized as follows. In the next section we recall the essential 
results of the density-matrix expansion of Negele and Vautherin \cite{negele} 
which provides the adequate technical framework to compute the nuclear energy 
density functional. In section 3, we then present analytical formulas for the 
various strength functions $F_\tau(k_f)$, $F_d(k_f)$, $F_{so}(k_f)$ and 
$F_J(k_f)$ entering the nuclear energy density functional. These expressions 
are derived (exclusively) from the two-loop $1\pi$-exchange Fock diagram and 
the three-loop iterated $1\pi$-exchange Hartree and Fock diagrams. Section 4 is
devoted to a discussion of our results and finally section 5 ends with a 
summary and an outlook. 

\section{Density-matrix expansion and energy density functional}
The starting point for the construction of an explicit nuclear energy density 
functional is the density-matrix as given by a sum over the occupied energy 
eigenfunctions $\Psi_\alpha$ of this many-fermion system. According to Negele 
and Vautherin \cite{negele} the bilocal density-matrix can be expanded in 
relative and center-of-mass coordinates, $\vec a$  and $\vec r$, as follows:
\begin{eqnarray} \sum_{\alpha\in occ}\Psi_\alpha( \vec r -\vec a/2)\Psi_\alpha^
\dagger(\vec r +\vec a/2) &=& {3 \rho\over a k_f}\, j_1(a k_f)-{35 \over 2a
k_f^3} \,j_3(a k_f) \bigg[ \tau - {3\over 5} \rho k_f^2 - {1\over 4} \vec
\nabla^2 \rho \bigg] \nonumber \\ && + {i \over 2} \,j_0(a k_f)\, \vec \sigma
\cdot (\vec a \times \vec J\,) + \dots\,,  \end{eqnarray}
where the functions $j_l(ak_f)$ are ordinary spherical Bessel functions. The
other quantities appearing on the right hand side of eq.(1) are the local 
nucleon density: 
\begin{equation} \rho(\vec r\,) =  {2k_f^3(\vec r\,)\over 3\pi^2} = \sum_{
\alpha\in occ}\Psi^\dagger_\alpha( \vec r\,)\Psi_\alpha( \vec r\,)
\,,\end{equation} 
written here in terms of the local Fermi-momentum $k_f(\vec r\,)$, the local 
kinetic energy density: 
\begin{equation} \tau(\vec r\,) = \sum_{\alpha\in occ}\vec \nabla\Psi^\dagger
_\alpha( \vec r\,)\cdot \vec \nabla\Psi_\alpha( \vec r\,) \,,\end{equation}
and the local spin-orbit density:
\begin{equation} \vec J(\vec r\,) = \sum_{\alpha\in occ} \Psi^\dagger_\alpha( 
\vec r\,)i\,\vec \sigma \times \vec \nabla\Psi_\alpha( \vec r\,) \,.
\end{equation}
For notational simplicity we have dropped their argument $\vec r$ in eq.(1) and
will do so in the following. It is important to note that a pairwise filling of
time-reversed orbitals $\alpha$ has been assumed in eq.(1). If the many-body 
ground state is not time-reversal invariant (as it is the case for odd nuclei 
and for rotating nuclei) various additional time-reversal-odd fields come into
play \cite{reinhard}. The local spin-orbit density $\vec J(\vec r\,)$ is 
non-zero for spin-unsaturated shells. In such a situation the density-matrix
is no longer a scalar in spin-space but has also a vector part. The Fourier 
transform of the density-matrix eq.(1) with respect to both coordinates $\vec 
a$ and  $\vec r$ defines the medium insertion for the inhomogeneous 
many-nucleon system characterized by the time-reversal-even fields $\rho(\vec 
r\,)$, $\tau(\vec r\,)$ and $\vec J(\vec r \,)$:
\begin{eqnarray} \Gamma(\vec p,\vec q\,)& =& \int d^3 r \, e^{-i \vec q \cdot
\vec r}\,\bigg\{ \theta(k_f-|\vec p\,|) \bigg[1 +{35 \pi^2 \over 8k_f^7}(5\vec
p\,^2 -3k_f^2) \bigg( \tau - {3\over 5} \rho k_f^2 - {1\over 4} \vec \nabla^2 
\rho \bigg) \bigg] \nonumber \\ && \quad \qquad \qquad +{\pi^2 \over 4k_f^4}
\Big[\delta(k_f-|\vec p\,|) -k_f \,\delta'(k_f-|\vec p\,|) \Big]\, \vec \sigma
\cdot (\vec p \times \vec J\,)  \bigg\}\,.  \end{eqnarray}
The double line in the left picture of Fig.\,1 symbolizes this medium insertion
together with the assignment of the out- and in-going nucleon momenta $\vec p
\pm \vec q/2$. The momentum transfer $\vec q$ is provided by the Fourier 
components of the inhomogeneous (matter) distributions: $\rho(\vec r\,)$, 
$\tau(\vec r\,)$ and $\vec J(\vec r\,)$. As a check one verifies that the 
Fourier transform $(1/2\pi^3) \int d^3 p\, e^{-i\vec p \cdot \vec a}$ of the 
(partly very singular) expression in the curly brackets in eq.(5) gives exactly
the right hand side of the density-matrix expansion in eq.(1). For homogeneous 
nuclear matter (where $\tau = 3 \rho k_f^2/5$ and $\vec \nabla \rho =\vec J=
\vec 0$) only the familiar step-function $\theta(k_f-|\vec p\,|)$ remains from
the medium insertion eq.(5) as the density of nucleon states in momentum space.

Going up to second order in spatial gradients (i.e. deviations from 
homogeneity) the energy density functional relevant for N=Z even-even nuclei 
reads \cite{ring}: 
\begin{eqnarray} {\cal E}[\rho,\tau,\vec J\,] &=& \rho\,\bar E(k_f)+\bigg[\tau-
{3\over 5} \rho k_f^2\bigg] \bigg[{1\over 2M}-{5k_f^2 \over 56 M^3}+F_\tau(k_f)
\bigg] \nonumber \\ && + (\vec \nabla \rho)^2\, F_\nabla(k_f)+  \vec \nabla 
\rho \cdot\vec J\, F_{so}(k_f)+ \vec J\,^2 \, F_J(k_f)\,.\end{eqnarray} 
Here, $\bar E(k_f)$ is the energy per particle of isospin symmetric nuclear
matter evaluated at the local Fermi momentum $k_f(\vec r\,)$. The (small) 
relativistic correction term $-5k_f^2/56M^3$ has been included in eq.(6) for 
the following reason. When multiplied with $-3\rho k_f^2/5$ it cancels together
with the foregoing term $1/2M$ the relativistically improved kinetic energy in
$\bar E(k_f)$ (see eq.(5) in ref.\cite{nucmat}). The functions $F_\tau(k_f)$, 
$F_\nabla(k_f)$, $F_{so}(k_f)$ and $F_J(k_f)$ arising from NN-interactions 
encode new dynamical information specific for inhomogeneous many-nucleon 
systems. In Skyrme parameterizations $F_\tau(k_f)$ depends linearly on the 
(local) density $\rho = 2k_f^3/3\pi^2$ whereas $F_{\nabla,so,J}(k_f)$ are just 
constants. Note that $F_{so}(k_f)$ gives the strength of the nuclear spin-orbit
coupling while $F_\nabla(k_f)$ is responsible for the formation of the nuclear 
surface. Variation of the energy density functional ${\cal E}[\rho,\tau,\vec 
J\,]$ with respect to single-particle wavefunctions under the condition that
these are normalized to unity leads to self-consistent density dependent
Hartree-Fock equations \cite{ring}. 

Returning to the medium insertion in eq.(5) one sees that the strength function
$F_\tau(k_f)$ emerges via a perturbation on top of the density of states 
$\theta(k_f-|\vec p\,|)$. The single particle potential in nuclear matter can 
actually be constructed in the same way by introducing a delta-function like 
perturbation \cite{pot}. Consequently, the strength function $F_\tau(k_f)$ can
be directly expressed in terms of the real part $U(p,k_f)$ of the momentum and density dependent single particle potential as:    
\begin{equation} F_\tau(k_f) = {35 \over 4k_f^7} \int_0^{k_f} dp\,
p^2(5p^2-3k_f^2)\, U(p,k_f) \,.\end{equation}
In eq.(5) the term $\tau-3\rho k_f^2/5$ is accompanied by $-\vec \nabla^2 
\rho/4$. Performing a partial integration of the energy $\int d^3r\,{\cal E}$
one sees immediately that part of the strength function $F_\nabla(k_f)$ is 
given by the $\rho$-derivative of $F_\tau(k_f)/4$. These considerations lead 
to the following decomposition:     
\begin{equation} F_\nabla(k_f) = {\pi^2 \over 8 k_f^2}\, {\partial F_\tau(k_f)
\over  \partial k_f} +F_d(k_f) \,,\end{equation}
where $F_d(k_f)$ comprises all those contributions for which the $(\vec \nabla
\rho)^2$-factor originates directly from the interactions. An example for this
mechanism will be explained in the next section.  

As a check on the present formalism (summarized in eq.(5)) we rederived the
Skyrme energy density functional (eq.(5.87) in ref.\cite{ring}) from the
matrix elements of the underlying two-body potential (eq.(4.105) in
ref.\cite{ring}) in a purely diagrammatic framework. In the next section we
use the same formalism to compute the nuclear energy density functional eq.(6) 
from one- and two-pion exchange diagrams.
\section{Diagrammatic calculation}
In this section we present analytical formulas for the four strength functions
$F_\tau(k_f)$, $F_d(k_f)$, $F_{so}(k_f)$ and $F_J(k_f)$ as derived from 
the two-loop one-pion exchange Fock diagram and the three-loop iterated 
one-pion exchange Hartree and Fock diagrams. These graphs are shown in Fig.\,1.
We give for each diagram only the final result omitting all technical details 
related to extensive algebraic manipulations and solving elementary integrals. 

\begin{center}
\SetWidth{2.5}
\begin{picture}(400,120)

\Line(20,0)(20,100)
\Line(13,47)(27,47)
\Line(13,53)(27,53)
\LongArrow(20,25)(20,30)
\LongArrow(20,70)(20,75)
\Text(-15,50)[]{$-\Gamma(\vec p,\vec q\,)$}
\Text(-10,15)[]{$\vec p-\vec q/2$}
\Text(-10,85)[]{$\vec p+\vec q/2$}
\Vertex(20,0){4}
\Vertex(20,100){4}
\Text(47,0)[]{$\vec r+\vec a/2$}
\Text(47,100)[]{$\vec r-\vec a/2$}

\ArrowArc(120,50)(40,0,180)
\ArrowArc(120,50)(40,180,360)
\DashLine(80,50)(160,50){6}
\Vertex(80,50){4}
\Vertex(160,50){4}

\CArc(230,50)(40,90,270)
\ArrowLine(230,10)(230,90)
\DashLine(230,90)(290,90){6}
\Vertex(230,90){4}
\Vertex(290,90){4}
\CArc(290,50)(40,-90,90)
\ArrowLine(290,10)(290,90)
\DashLine(230,10)(290,10){6}
\Vertex(230,10){4}
\Vertex(290,10){4}

\ArrowArc(400,50)(40,0,180)
\ArrowArc(400,50)(40,180,360)
\DashLine(428.3,78.3)(371.7,21.7){6}
\DashLine(428.3,21.7)(371.7,78.3){6}
\Vertex(428.3,78.3){4}
\Vertex(371.7,21.7){4}
\Vertex(428.3,21.7){4}
\Vertex(371.7,78.3){4}

\end{picture}
\end{center}
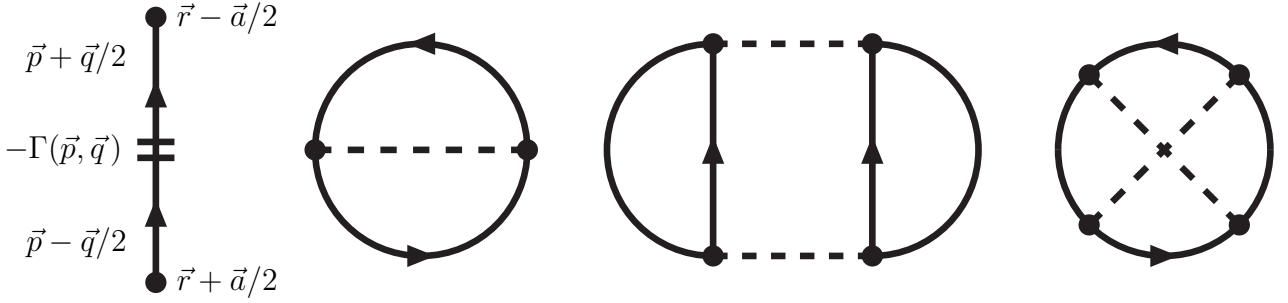
{\it Fig.\,1: Left: The double line symbolizes the medium insertion defined by
eq.(5). Next are shown: The two-loop one-pion exchange Fock-diagram and the
three-loop iterated one-pion exchange Hartree- and Fock-diagrams. The
combinatoric factors of these diagrams are 1/2, 1/4 and 1/4, in the order 
shown. Their isospin factors for isospin symmetric nuclear matter are 6, 12 and
$-$6, respectively.}

\subsection{One-pion exchange Fock diagram with two medium insertions}
The non-vanishing contributions read:
\begin{equation} F_\tau(k_f) = {35g_A^2m_\pi\over(16\pi f_\pi)^2 u^5} 
\bigg\{{4\over 3}u^4 +24u^2-1 -20u \arctan 2u +\bigg( {9\over 2}-6u^2 +{1\over 
4u^2}\bigg) \ln(1+4u^2) \bigg\} \,, \end{equation}
\begin{equation} F_J(k_f) = {g_A^2\over(8 m_\pi f_\pi)^2} \bigg\{{10+24u^2 
\over (1+4u^2)^2}+{1\over 2u^2} \ln(1+4u^2) \bigg\} \,, \end{equation}
with $u=k_f/m_\pi$ the ratio of the two small scales inherent to our
calculation. The expression for $F_\tau(k_f)$ in eq.(9) follows simply from
inserting the static $1\pi$-exchange single particle potential (eq.(8) in
ref.\cite{pot} in the limit $M\to \infty$) into the "master formula" eq.(7). 
The vanishing of $F_d(k_f)$ has the following reason. The momentum transfer 
$\pm \vec q$ at the upper and lower medium insertion does not flow into the 
exchanged virtual pion line (because of momentum conservation at each 
interaction vertex). Therefore there is no factor of $\vec q\,^2$ which
could produce via Fourier transformations a $(\vec \nabla \rho)^2$-factor. 
The spin-orbit strength $F_{so}(k_f)$ vanishes as a result of the spin-trace: 
tr$[\vec \sigma \cdot (\vec p_1-\vec p_2)\, \vec \sigma \cdot (\vec p_{1,2} 
\times \vec J\,)\,\vec \sigma \cdot (\vec p_1-\vec p_2)] = 0$.
 
\subsection{Iterated one-pion exchange Hartree diagram with two medium
insertions} 
We find the following closed form expressions:
\begin{eqnarray} F_\tau(k_f) &=& {g_A^4M m_\pi^2\over(8\pi)^3 (u f_\pi)^4} 
\bigg\{{151\over 3}u^3 -(350+16u^4) \arctan 2u \nonumber \\ && +444u-{55\over 
4u}+\bigg( {55\over 16u^3}+{567 \over 8u}-{245\over 2}u \bigg) \ln(1+4u^2)
\bigg\} \,, \end{eqnarray}
\begin{equation} F_d(k_f) = {g_A^4M \over \pi m_\pi(4f_\pi)^4} \bigg\{
{4\over u} \arctan 2u -{23 \over 16u^2} \ln(1+4u^2) -{3+20u^2\over 12(1+4u^2
)^2}  \bigg\} \,,  \end{equation}
\begin{equation} F_{so}(k_f) = {g_A^4M \over \pi m_\pi(4f_\pi)^4} \bigg\{
{4\over 1+4u^2} -{3 \over 2u^2} \ln(1+4u^2)\bigg\} \,.  \end{equation}
Again, $F_\tau(k_f)$ in eq.(11) stems from inserting the two-body potential
$U_2(p,k_f)$ (eq.(9) in ref.\cite{pot}) into the "master formula" eq.(7). Note
that any $p$-independent contribution, in particular the cut-off dependent term
eq.(17) in ref.\cite{pot}, drops out. The vanishing of $F_J(k_f)$ results from 
the spin-trace over a nucleon ring being equal to zero (as demonstrated at the
end of section 3.1). Let us briefly explain the mechanism which generates the 
strength function $F_d(k_f)$. The exchanged pion-pair transfers the momentum 
$\vec q$ between the left and the right nucleon ring. This momentum $\vec q$ 
enters both the pseudovector $\pi N$-interaction vertices and the pion 
propagators. After expanding the inner loop integral to order $\vec q\,^2$ the 
Fourier transformation in eq.(5) converts this factor $\vec q\,^2$ into a 
factor $(\vec \nabla k_f)^2$. The rest is a solvable integral over the product 
of two Fermi surfaces. The spin-orbit strength $F_{so}(k_f)$ arises from the 
spin-trace: tr$[\vec \sigma \cdot (\vec l+\vec q/2)\,\vec \sigma\cdot (\vec l
-\vec q/2)\,\vec \sigma \cdot (\vec p_{1,2} \times \vec J\,)] = 2i\,(\vec q 
\times \vec l\,)\cdot(\vec p_{1,2} \times \vec J\,)$ where $\vec q$ gets again 
converted to $\vec \nabla k_f$ by Fourier transformation. The remainder is a 
solvable integral over delta-functions and derivatives thereof. 

\subsection{Iterated one-pion exchange Fock diagram with two medium insertions}
We find the following contributions from the last diagram in Fig.\,1 with two
medium insertions at non-neighboring nucleon propagators:
\begin{eqnarray} F_\tau(k_f) &=& {35g_A^4 M m_\pi^2 \over (4\pi)^3f_\pi^4
u^7}\int_0^u \!dx {x^2 (u-x)^2\over 2(1+2x^2)}(2x^2+4ux-3u^2)\nonumber \\&& 
\times \Big[(1+8x^2+8x^4) \arctan x-(1+4x^2)\arctan2x\Big] \,, \end{eqnarray}
\begin{eqnarray} F_d(k_f) &=& {g_A^4M \over \pi m_\pi(8f_\pi)^4} \bigg\{
{4\over u}( \arctan u-2 \arctan 2u)+{1\over u^2}\ln{(1+2u^2)(1+4u^2)\over 
(1+u^2)^2} \nonumber \\ &&+{4\over 1+2u^2} +{2\over u^2}\int_0^u \!dx
{3+18x^2+16x^4\over (1+2x^2)^3}\Big[\arctan 2x-\arctan x\Big] \bigg\} \,,  
\end{eqnarray}
\begin{eqnarray} F_{so}(k_f) &=& {g_A^4M \over \pi m_\pi(4f_\pi)^4} \bigg\{
{1\over 4u^2}\ln{1+4u^2\over 1+u^2} +{3+4u^2 \over 2u(1+2u^2)}\arctan u 
\nonumber \\ && -{\arctan 2u \over 2u(1+2u^2)} +{1\over 2u^2}\int_0^u \!dx
{\arctan x-\arctan 2x\over 1+2x^2} \bigg\} \,,  \end{eqnarray}
\begin{eqnarray} F_J(k_f) &=& {g_A^4M \over \pi m_\pi(8f_\pi)^4} \bigg\{
{2 \arctan 2u \over u(1+2u^2)}-{2(5+8u^2)\over u(1+2u^2)}\arctan u  -{1\over 
1+u^2}\nonumber \\ &&-{1\over u^2}\ln{1+4u^2\over 1+u^2}+{2\over u^2}\int_0^u
\!dx {\arctan 2x-\arctan x\over 1+2x^2} \bigg\} \,.  \end{eqnarray}
The basic mechanisms which lead to these results are the same as explained
before. Concerning kinematics and spin-algebra the iterated $1\pi$-exchange
Fock diagram is somewhat more involved than the Hartree diagram. Even though 
all occurring inner $d^3l$-loop integrals can be solved in closed form there 
remain some non-elementary integrals from the integration over the product of
two Fermi spheres of radius $k_f$. 

\subsection{Iterated one-pion exchange Hartree diagram with three medium
insertions} 
In this case we find the following contributions:
\begin{eqnarray} F_\tau(k_f) &=& {175g_A^4 M m_\pi^2 \over (4\pi f_\pi)^4 u^7}
\int_0^u \!dx x^2\int_{-1}^1 \!dy \bigg\{ u^3 x y \bigg[\ln(1+s^2)-{2s^2+s^4
\over 2(1+s^2)}\bigg] + \Big[2uxy+(u^2-x^2y^2)H\Big] \nonumber \\ && \times 
\bigg[ {3\over 4}\Big(3+{13\over 5}u^2-4x^2-x^2y^2\Big) \ln(1+s^2) +\Big(4x^2
+x^2y^2 -{13\over 5}u^2-2\Big){3(2s^2+s^4)\over 8(1+s^2)} \nonumber \\ &&
+{3\over 8}s^2(s^2-2)+s x y(6-s^2) +{3s x y \over 2(1+s^2)} -{15 \over 2} x y
\arctan s \bigg]\bigg\}\,,  \end{eqnarray}
with the auxiliary functions $H=\ln(u+xy)-\ln(u-xy)$ and $s=xy +\sqrt{u^2-x^2+
x^2y^2}$. The quantity $s$ has the geometrical meaning of the distance between 
a point on a sphere of radius $u$ and an interior point displaced at a distance
$x$ from the center of the sphere. In the same geometrical picture $y$ denotes 
a directional cosine.  
\begin{eqnarray} F_d(k_f) &=& {g_A^4 M u^{-4}\over \pi^2 m_\pi(4  f_\pi)^4}
\int_0^u \!dx x^2\int_{-1}^1 \!dy \bigg\{ H\bigg[{s^2\over 4(1+s^2)^4}(7s^6+38
s^4 +63s^2+24)\nonumber \\ && -6\ln(1+s^2)\bigg] + {uxy \over u^2-x^2y^2} 
\bigg[ {s^2\over 6(1+s^2)^3}(23s^4 +51s^2+24)-4\ln(1+s^2)\bigg]\bigg\}\,,  
\end{eqnarray}
\begin{eqnarray} F_{so}(k_f) &=& {2g_A^4 M u^{-6}\over \pi^2 m_\pi(4  f_\pi)^4}
\int_0^u \!dx x^2\int_{-1}^1 \!dy \bigg\{{2u s^4 [2 x^2y^2(s-s')-u^2 s] \over 
(1+s^2)^2 (u^2-x^2y^2)}\nonumber \\ && + \bigg[{u(3u^2-5x^2y^2)\over u^2-
x^2y^2} -4xy H \bigg] \bigg[ 3 \arctan s-{3s+2s^3\over 1+s^2}\bigg] + {H s^4  
\over (1+s^2)^3}\nonumber \\ && \times \Big[(5+s^2)s'^2-2xys'(7+3s^2)+sxy 
(11+7s^2)+(s+s^3)(s''-s') \Big]\bigg\}\,.   \end{eqnarray}
Here we have introduced the partial derivatives $s'=u \partial s/\partial u$
and $s''=u^2 \partial^2 s/\partial u^2$. 
\begin{eqnarray} F_J(k_f) &=& {g_A^4 M \over \pi^2 m_\pi(4  f_\pi)^4} \bigg\{
{96u^6+24u^4-12u^2-1 \over u(1+4u^2)^3} +{1+2u^2 \over 4u^3} \ln(1+4u^2) 
+\int_0^1 \!dy\, {8u^3 y^2 \over (1+4u^2y^2)^4} \nonumber \\ && \times 
\Big[ (30+32u^2)y^2-5+(16u^4-24u^2-35)y^4-56u^2y^6-48 u^4
y^8\Big]\ln{1+y \over 1-y} \bigg\} \,. \end{eqnarray}
The last contribution $F_J(k_f)$ in eq.(21) is obtained when both insertions
proportional to $\vec \sigma \cdot (\vec p_{1,2}\times \vec J\,)$ (producing,
after integration, the overall $\vec J\,^2$-factor) are under a single
spin-trace. For the other two possible combinations the spin-traces are equal
to zero.  

\subsection{Iterated one-pion exchange Fock diagram with three medium
insertions}
The evaluation of this diagram is most tedious. It is advisable to split the
contributions to the four strength functions $F_\tau(k_f)$, $F_d(k_f)$, 
$F_{so}(k_f)$ and $F_J(k_f)$ into "factorizable" and "non-factorizable" parts. 
These two pieces are distinguished by whether the nucleon propagator in the 
denominator can be canceled or not by terms from the product of $\pi N
$-interaction vertices in the numerator. We find the following "factorizable" 
contributions: 
\begin{eqnarray} F_\tau(k_f) &=& {35g_A^4 M m_\pi^2 \over 2(8\pi f_\pi)^4 u^7} 
\int_0^u  \!dx\, \Big[u(1+u^2+x^2)-[1+(u+x)^2][1+(u-x)^2]L \Big]\nonumber \\
&&\times \bigg\{ 5u (5x^4+19x^2-2) -u^3 \Big({26\over 3}x^2+17\Big)
-7u^5 \nonumber \\ && -80x^2 \Big[ \arctan(u+x)+\arctan(u-x)\Big] +
\Big[(1+u^2)^2(10+7u^2)\nonumber \\ &&+3x^2(25+8u^2-13u^4)+3x^4
(19u^2-40)-25x^6 \Big] L\bigg\}\,,\end{eqnarray}
with the auxiliary function:
\begin{equation} L= {1\over 4x} \ln{1+(u+x)^2\over 1+(u-x)^2} \,,\end{equation}
\begin{eqnarray} F_d(k_f) &=& {g_A^4 M u^{-2} \over \pi^2 m_\pi (8f_\pi)^4} 
\Bigg\{ {1+6u^2 \over 4u^3} \ln^2(1+4u^2)-{5+52u^2+104u^4\over 2u(1+u^2)
(1+4u^2)} \ln(1+4u^2) \nonumber \\ && -{19u^2 \over 1+u^2} \arctan 2u
+{6u(1+8u^2) \over 1+4u^2}+8 \int_0^u \!dx\,\bigg\{\Big[1+u^2+3(1+u^2)^2
x^{-2}\Big] L^2 \nonumber \\ && + 3u^2 x^{-2} +\Big[(4u-x)[1+(u+x)^2]^{-1}
+ (4u+x)[1+(u-x)^2]^{-1}\nonumber \\ &&-4u-6(u+u^3)x^{-2}-2x[1+(u+x)^2 ]^{-2}
+ 2x[1+(u-x)^2]^{-2} \Big]L\bigg\} \Bigg\} \,,\end{eqnarray}
\begin{eqnarray} F_{so}(k_f) &=& {g_A^4 M u^{-3} \over \pi^2 m_\pi (4f_\pi)^4} 
\Bigg\{ {1+2u^2\over 32u^2} \ln^2(1+4u^2)-{3u^4 \ln(1+4u^2) \over (1+u^2)
(1+4u^2)} +{u^2(3+20u^2) \over 2(1+4u^2)} \nonumber \\ && -{u(1+11u^2+16u^4) 
\over (1+u^2)(1+4u^2)} \arctan 2u + \int_0^u \!dx\,\bigg\{\Big[3(1+u^2)^2 
x^{-2}-4x^2-1-u^2\Big]u L^2 \nonumber \\ && +\Big[(1+5u^2+5ux)[1+(u+x)^2]^{-1}
+ (1+5u^2-5ux)[1+(u-x)^2]^{-1}\nonumber \\ && -6(u^2+u^4)x^{-2} -2\Big]L+
3u^3x^{-2}\bigg\}  \Bigg\} \,,\end{eqnarray}
\begin{eqnarray} F_J(k_f) &=& {g_A^4 M u^{-3} \over \pi^2 m_\pi (8f_\pi)^4} 
\Bigg\{ {3+12u^2+8u^4 \over 4u^4} \ln^2(1+4u^2)-{7+8u^2+8u^4\over u^2(1+u^2)} 
\ln(1+4u^2) \nonumber \\ && +{2(5-2u^2)(2+3u^2) \over u(1+u^2)} 
\arctan 2u +{8(8u^6-158u^4-73u^2-9) \over 3(1+4u^2)^2}\nonumber \\ && +8 
\int_0^u \!dx\,\bigg\{\Big[3(1+u^2)^2 x^{-2}+3x^2-2-2u^2\Big]u L^2 
+2\Big[2+u^2-3x^{-2}(u^2+u^4)\\ && +(u^2-ux-1)[1+(u+x)^2]^{-1} + (u^2+ux-1)
[1+(u-x)^2]^{-1}\Big]L+ 3u^3x^{-2}\bigg\} \Bigg\} \nonumber \,. \end{eqnarray}
The "non-factorizable" contributions read on the other hand:
\begin{eqnarray} F_\tau(k_f) &=& {35g_A^4 M m_\pi^2 \over (8\pi f_\pi)^4 u^7}
\int_0^u \!dx \,x^2\int_{-1}^1\!dy \int_{-1}^1 \!dz {yz \,\theta(y^2+z^2-1) 
\over |yz|\sqrt{y^2+z^2-1}}\nonumber \\ && \times \Big[t^2- \ln(1+t^2)\Big]
\bigg\{ (45x^2-27u^2-30) \ln(1+s^2)\nonumber \\&& +120 x y \arctan s +2s x y 
(17u^2-30-35x^2+20x^2y^2) \bigg\}\,, \end{eqnarray}
\begin{eqnarray} F_d(k_f) &=& {g_A^4 M u^{-6}\over \pi^2 m_\pi(8f_\pi)^4}
\int_0^u \!dx \,x^2\int_{-1}^1\!dy \int_{-1}^1 \!dz {yz \,\theta(y^2+z^2-1) 
\over |yz|\sqrt{y^2+z^2-1}}\bigg\{ \bigg[2\ln(1+t^2)- {t^2(3+t^2) \over
(1+t^2)^2} \bigg]\nonumber \\ && \times {2s^2 \over (1+s^2)^2} 
\Big[(6s+4s^3)s'-(3+s^2)s'^2-(s+s^3)s''\Big] +{4s^3s' t^2(2t^4+5t^2-1) \over
(1+s^2)(1+t^2)^3} \bigg\}\,, \end{eqnarray}
\begin{eqnarray} F_{so}(k_f) &=& {g_A^4 M \over \pi^2 m_\pi(4f_\pi)^4}
\int_{-1}^1\!dy \int_{-1}^1 \!dz {yz \,\theta(y^2+z^2-1)  \over |yz|\sqrt{y^2+
z^2-1}}\bigg\{ 16y^2z\,\theta(y)\theta(z) \bigg[{1+2u^2y^2 \over (1+4u^2y^2)^2}
\nonumber \\ && \times \Big(\arctan 2uz-2uz\Big) +{ u^3z(1-2z^2)\over
(1+4u^2y^2)(1+4u^2 z^2)} \bigg]+\int_0^u \!dx \, {u^{-8}x^2 st^2 t'\over 2 
(1+s^2)^2(1+t^2)} \nonumber \\ && \times \Big[(s+s^3) (s'-s'')(st+sxz -txy) 
+s'^2(2txy -(3s+s^3)(t+xz) ) \Big]\bigg\}\,, \end{eqnarray} 
\begin{eqnarray} F_J(k_f) &=& {g_A^4 M \over \pi^2 m_\pi(4f_\pi)^4}
\int_{-1}^1\!dy \int_{-1}^1 \!dz {yz \,\theta(y^2+z^2-1) 
\over |yz|\sqrt{y^2+z^2-1}}\bigg\{ y^2\, \theta(y)\theta(z) \bigg[
\Big[\ln(1+4u^2 z^2)-4u^2 z^2 \Big] \nonumber \\ && \times {9 +4u^2(5+2y^2)
+16u^4 (y^2+y^4) \over u(1+4u^2y^2)^3} +{ 16u^3(3+4u^2y^2)z^2 (1-2z^2) \over  
(1+4u^2y^2)^2(1+4u^2 z^2)} \bigg]\nonumber \\ &&+\int_0^u \!dx \,
{x^4 s^2t^2(1-y^2-z^2) \over 4u^{10} (1+s^2)^2(1+t^2)^2}\Big[(s+s^3)(s''-s')+
(3+s^2)s'^2\Big]\nonumber \\ && \times \Big[(t+t^3)(t''-t')+ (3+t^2)t'^2\Big]
\bigg\}\,, \end{eqnarray}
with the auxiliary function $t=xz +\sqrt{u^2-x^2+x^2z^2}$ and its partial
derivatives $t'=u \partial t/\partial u$ and $t''=u^2 \partial^2 t/\partial 
u^2$. For the numerical evaluation of the $dydz$-double integrals in
eqs.(27-30) it is advantageous to first antisymmetrize the integrands both in 
$y$ and $z$ and then to substitute $z = \sqrt{y^2 \zeta^2+1-y^2}$. This way the
integration region becomes equal to the unit-square $0< y,\zeta <1$. 

Obviously, the analytical results presented in this section do not involve any
adjustable parameter. Only well-known physical quantities like the nucleon 
axial-vector coupling constant $g_A = 1.3$, the nucleon mass $M=939\,$MeV, the
pion decay constant $f_\pi = 92.4\,$MeV and the (neutral) pion mass $m_\pi =
135\,$MeV enter.  

Let us end this section with general power counting considerations for the 
nuclear energy density functional ${\cal E}[\rho,\tau,\vec J\,]$. Counting the 
Fermi momentum $k_f$, the pion mass $m_\pi$ and a spatial gradient $\vec
\nabla$ collectively as small momenta one deduces from eqs.(2,3,4) that the
nucleon density $\rho(\vec r\,)$, the kinetic energy density $\tau(\vec r\,)$
and the spin-orbit density  $\vec J(\vec r\,)$ are quantities of third, fifth
and fourth order in small momenta, respectively. With these counting rules the
contributions from $1\pi$-exchange to the nuclear energy density functional
${\cal E}[\rho,\tau,\vec J\,]$ are of sixth order in small momenta while all
contributions from iterated $1\pi$-exchange are of seventh order. Concerning
NN-interactions induced by pion-exchange the nuclear energy density functional 
presented here is in fact complete up-to-and-including seventh order in small 
momenta.  
  
\section{Results and discussion}
In this section we present and discuss our numerical results using the input
parameters just mentioned. Returning to the energy density functional ${\cal
E}[\rho,\tau,\vec J\,]$ in eq.(6) one observes that the expression in square 
brackets multiplying the kinetic energy density $\tau(\vec r\,)$ has the 
interpretation of a reciprocal density dependent effective nucleon mass:
\begin{equation} \widetilde M^*(\rho) = M \bigg[1-{5k_f^2 \over 28M^2}+ 2M\, 
F_\tau(k_f)\bigg]^{-1} \,. \end{equation}
We note as an aside that this effective nucleon mass $\widetilde M^*(\rho)$ is 
conceptually different from the so-called "Landau" mass which derives from the 
slope of the single particle potential $U(p,k_f)$ at the Fermi surface $p=k_f$.
Only if the (real) single particle potential has a simple quadratic dependence
on the nucleon momentum, $U(p,k_f) = U_0(k_f)+ U_1(k_f)\,p^2$, these two 
variants of effective nucleon mass agree with each other. 

\bigskip

\bild{msm.eps}{16}
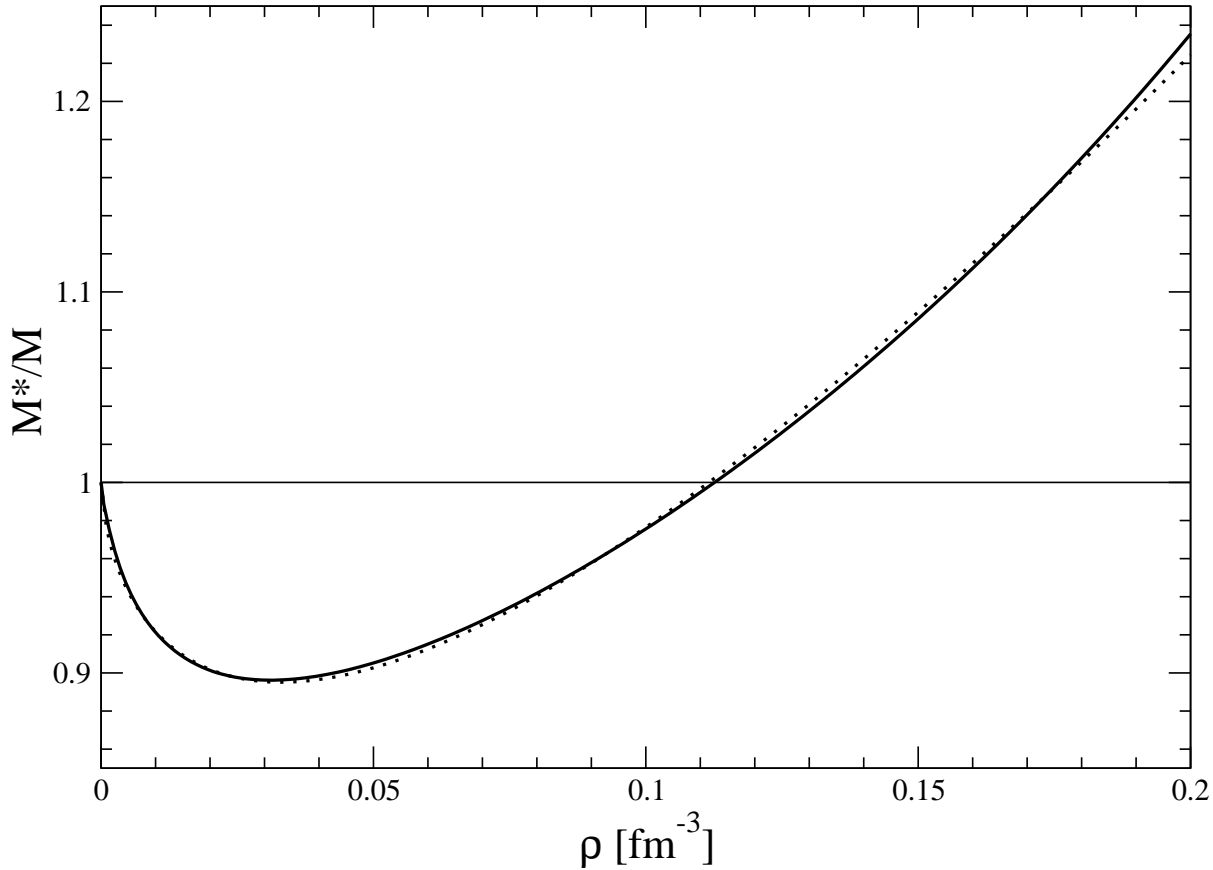
{\it Fig.\,2: The effective nucleon mass $\widetilde M^*(\rho)$ divided by the
free nucleon mass $M$ versus the nucleon density $\rho$. The dotted line
corresponds to the fit: $\widetilde M^*(\rho)/M= 1 - 3.054\, {\rm fm}^2\cdot
\rho^{2/3}+ 6.345 \,{\rm fm}^3\cdot\rho$.} 
\bigskip

In Fig.\,2 we show the ratio effective over free nucleon mass $\widetilde 
M^*(\rho)/M$ as a function of the nucleon density $\rho =2k_f^3/3\pi^2$. One
observes a reduced effective nucleon mass $0.89 M< \widetilde M^*(\rho)<M$ for 
densities $\rho<0.11\,{\rm fm}^{-3}$ and an enhanced effective nucleon mass 
$\widetilde M^*(\rho)>M$ for higher densities. In the region below the nuclear
matter saturation density $\rho < \rho_0 = 0.174\,{\rm fm}^{-3}$ relevant
for nuclear structure the deviations of the effective nucleon mass $\widetilde 
M^*(\rho)$ from its free space value $M$ do no exceed $\pm 15\%$. Let us give a
qualitative explanation for the (somewhat unusual) behavior of the curve in 
Fig.\,2. Consider the iterated $1\pi$-exchange Hartree diagram in Fig.\,1 at
sufficiently high densities such that the pion mass $m_\pi$ can be neglected
against the Fermi momentum $k_f$. In this (limiting) case the $\pi 
N$-interaction vertices get cancelled by the pion propagators. One is
effectively dealing with a zero-range contact interaction in second order which
according to Galitskii's calculation from 1958 \cite{galitski,fetter} generates
an enhanced in-medium mass. In this sense the curve in Fig.\,2 delineates the 
two density regimes $k_f < \sqrt{3}m_\pi$ and $k_f > \sqrt{3}m_\pi$ where the 
(qualitative) behavior in the latter is ruled by Galitskii's second order
result. Interestingly, a recent large scale fit of 1888 nuclide masses by 
Pearson et al. \cite{pearson} using a "Hartree-Fock nuclear mass formula" has 
given an effective nucleon mass of $\widetilde M^*(\rho_0)=1.05M$. This value 
is comparable with our parameterfree result $\widetilde M^*(\rho_0)=1.15M$. 

\bigskip

\bild{fgrad.eps}{16}
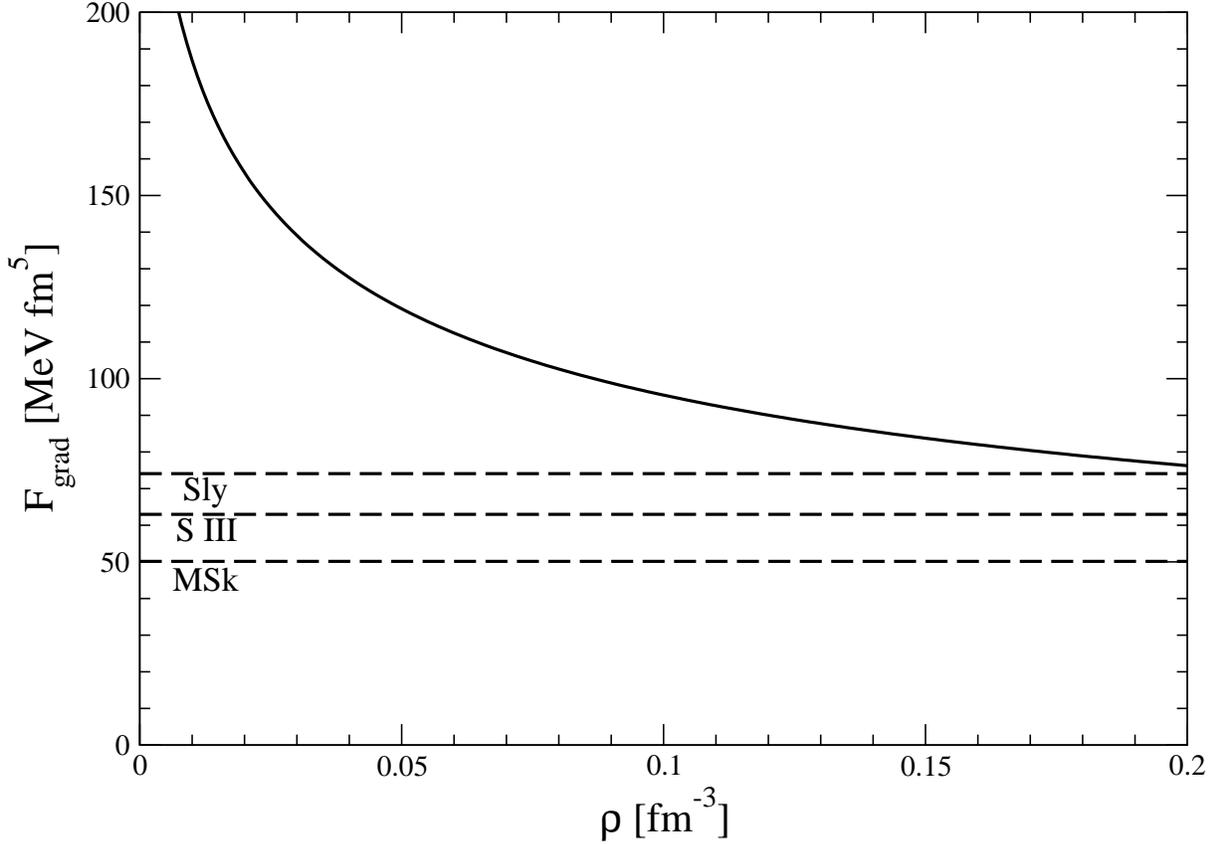
{\it Fig.\,3: The strength function $F_\nabla(k_f)$ related to the $(\vec
\nabla \rho)^2$-term in the nuclear energy density functional versus the 
nucleon density $\rho=2k_f^3/3\pi^2$. An accurate fit of the full line is: 
$F_\nabla(k_f)=  45.43\,{\rm MeVfm}^4\cdot\rho^{-1/3} -0.229\,{\rm MeVfm}^2
\cdot\rho^{-1}$. The three horizontal dashed lines show the constant values 
$F_\nabla(k_f)=[9t_1-(5+4x_2)t_2]/64$ of the Skyrme forces Sly \cite{sly},
 SIII \cite{sk3} and MSk \cite{pearson}.} 

\bigskip
The dotted line in Fig.\,2 corresponds to the fit: $\widetilde M^*(\rho)/M=1 - 
3.054\, {\rm fm}^2\cdot\rho^{2/3}+ 6.345 \,{\rm fm}^3\cdot\rho$, which may be
useful for applications in nuclear structure calculations. In this context we
mention also the fitted form of the underlying nuclear matter equation of state
\cite{lutzpot}: $\bar E(k_f) = 111.63\,{\rm MeVfm}^2\cdot \rho^{2/3} - 752.82
\, {\rm MeVfm}^3\cdot \rho +  832.74\,{\rm MeVfm}^4 \cdot \rho^{4/3}$. 

Next, we show in Fig.\,3 by the full line the strength function $F_\nabla(k_f)$
belonging to the $(\vec \nabla \rho)^2$-term in the nuclear energy density 
functional eq.(6) versus the nucleon density $\rho=2k_f^3/3\pi^2$. The three 
horizontal dashed lines represent the constant values $F_\nabla(k_f)=[9t_1-(5+
4x_2)t_2]/64$ of the Skyrme forces Sly \cite{sly}, SIII \cite{sk3} and MSk 
\cite{pearson}. In the case of Sly and MSk we have performed averages over the 
various parameter sets Sly4-7 and MSk1-6. At nuclear matter saturation density 
$\rho_0= 0.174\,{\rm fm}^{-3}$ our parameterfree prediction $F_\nabla(k_{f0})= 
80.1\,{\rm MeVfm}^5$ is comparable to these empirical values. The strong 
increase of the strength function $F_\nabla(k_f)$ with decreasing density has 
to do with the presence of a small mass scale, $m_\pi =135\,$MeV, and with
associated chiral singularities (of the form $m_\pi^{-2}$ and $m_\pi^{-1}$). We
will come back to this issue again towards the end of this section. An accurate
fit of the full line in Fig.\,3 is: $F_\nabla(k_f)=  45.43\,{\rm MeVfm}^4\cdot
\rho^{-1/3} -0.229\,{\rm MeVfm}^2 \cdot\rho^{-1}$. It may be useful for 
applications in nuclear structure calculations. Note also that the relevant 
contribution to the central single-particle potential, $-(\vec \nabla \rho)^2\,
\partial F_\nabla (k_f)/\partial \rho-2\vec \nabla^2  \rho\, F_\nabla (k_f)$, 
receives only little weight from very low densities. Therefore the deviation of
the strength function $F_\nabla(k_f)$ from a constant may be less dramatic in 
practice than it appears on first sight from Fig.\,3.

\bigskip

\bild{fso.eps}{16}
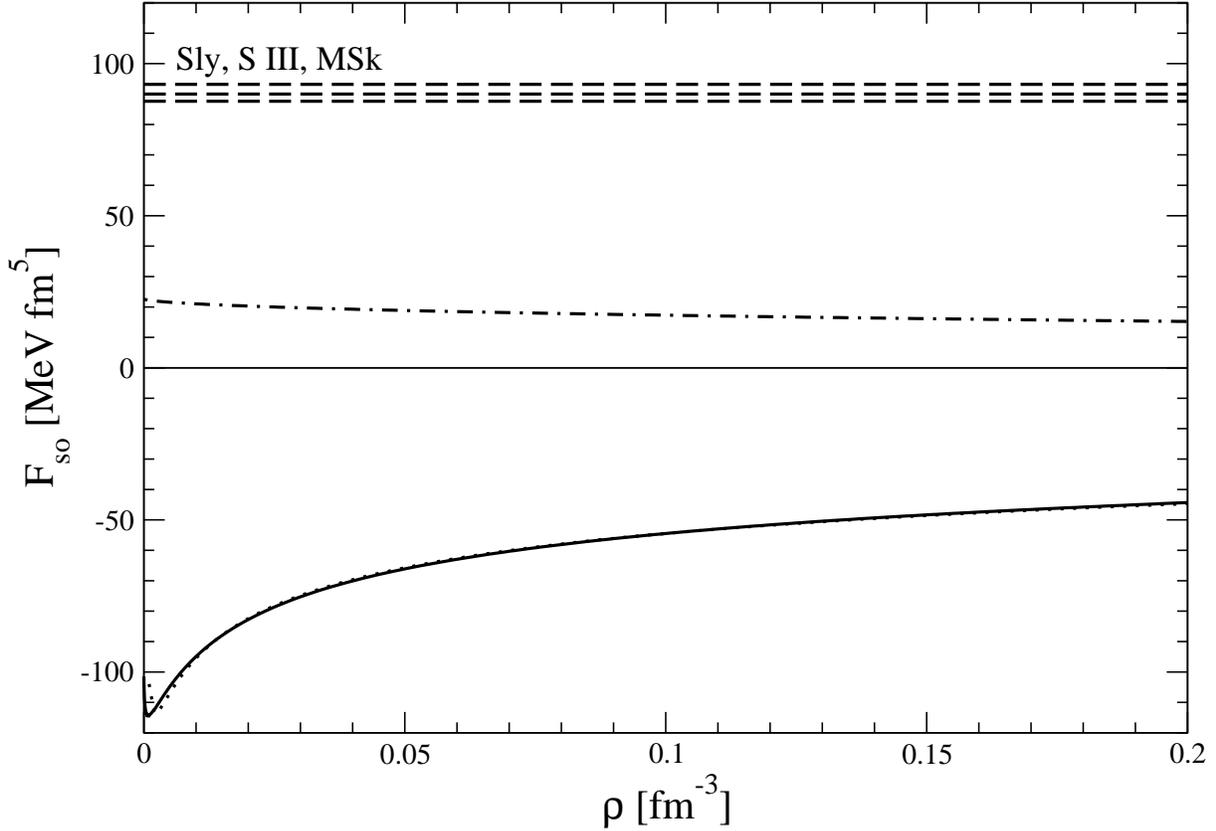
{\it Fig.\,4: The strength function $F_{so}(k_f)$ related to the spin-orbit
coupling term in the nuclear energy density functional versus the 
nucleon density $\rho=2k_f^3/3\pi^2$. The dotted line corresponds to the fit: 
$F_{so}(k_f)=  1.898\,{\rm MeVfm}^3\cdot\rho^{-2/3} - 29.37\,{\rm MeVfm}^4\cdot
\rho^{-1/3}$. The three horizontal dashed lines show the constant values 
$F_{so}(k_f)=3W_0/4$ of the Skyrme forces Sly \cite{sly}, SIII \cite{sk3} and 
MSk \cite{pearson}. The dashed-dotted line shows the contribution from
irreducible $2\pi$-exchange written in eq.(33) for a cut-off $\Lambda =
0.65\,$GeV.} 

\bigskip

The full line in Fig.\,4 shows the result of iterated $1\pi$-exchange for the 
strength function $F_{so}(k_f)$ belonging to the spin-orbit coupling term in 
the nuclear energy density functional. For comparison we have drawn the 
constant values $F_{so}(k_f)=3W_0/4$ of the three Skyrme forces Sly \cite{sly},
SIII \cite{sk3} and MSk \cite{pearson} (horizontal dashed lines). One observes 
that the strength of the nuclear spin-orbit interaction as generated by 
iterated $1\pi$-exchange is at $\rho_0$ about half as large as the 
corresponding empirical value, however, with the wrong negative sign. This 
"negative" result is dominated by the contribution of the iterated $1\pi
$-exchange Hartree diagram with two medium insertions (see eq.(13)). For 
example, one obtains numerically from eq.(13) at saturation density $\rho_0=
0.174\,{\rm fm}^{-3}$ (where $u=k_{f0}/m_\pi =2.0$) the negative value $F_{so}
(k_{f0}) = - 83.7\,{\rm MeVfm}^5$. The other diagrams with lower spin- and 
isospin weight factors reduce this number approximately half in magnitude. The 
"negative" result for $F_{so}(k_f)$ is to some extent already indicated by the 
calculation of the momentum and density dependent nuclear spin-orbit strength 
$U_{ls}(p,k_f)$ in ref.\cite{uls}. Going back to the medium insertion in eq.(5)
one learns that only the values of $U_{ls}(p,k_f)$ near the Fermi surface $p=
k_f$ will contribute to $F_{so}(k_f)$. As a matter of fact the curves in 
Fig.\,7 of ref.\cite{uls} drop from positive to negative values when $p$ runs 
from zero to $k_{f0}=272.7\,$MeV. Actually, for the contributions to $F_{so}
(k_f)$ from diagrams with two medium insertions eqs.(13,16) the following 
relationship holds: 
\begin{equation} F_{so}(k_f) = {\pi^2\over 4k_f^2} \bigg[ {\partial
U_{ls}(p,k_f) \over \partial k_f} + {k_f \over 3} \,{\partial^2 U_{ls}(p,k_f) 
\over  \partial p \partial k_f} \bigg]_{p=k_f} \,,\end{equation}
to be applied to the expressions $U_{ls}^{(a,e)}(p,k_f)$ in eqs.(9,17) of
ref.\cite{uls}. 

It is well-known that irreducible two-pion exchange generates (via relativistic
$1/M$-corrections) spin-orbit amplitudes in the T-matrix of elastic
nucleon-nucleon scattering \cite{nnpap}. Their effect on the nuclear spin-orbit
interaction has been calculated in ref.\cite{irred}. Inserting the expression 
$U_{ls}^{(2\pi)}(p,k_f)$ in eq.(18) of ref.\cite{irred} into the "master 
formula" eq.(32) one derives the following contribution from irreducible 
$2\pi$-exchange to the spin-orbit strength function: 
\begin{eqnarray} F_{so}(k_f) &=& {g_A^2 \over \pi M (4f_\pi)^4} \bigg\{
(16+19g_A^2) {\Lambda \over 2\pi}+ {m_\pi^3 \over 6 k_f^2}(4-3g_A^2) \ln{k_f^2
+m_\pi^2 \over m_\pi^2} \nonumber \\&& - {m_\pi \over 3}(8+27g_A^2)+{2\over 
3k_f} \Big[ 3m_\pi^2(g_A^2-2)-4k_f^2 \Big] \arctan{k_f \over m_\pi} \bigg\}
\,. \end{eqnarray}
Here, $\Lambda$ is a momentum cut-off which has been used to regularize the 
linear divergences of the irreducible $2\pi$-exchange (triangle and box)
diagrams. In dimensional regularization (employed in eqs.(22,23) of 
ref.\cite{nnpap}) such linear divergences are not visible. The dashed-dotted
line in Fig.\,4 shows the relatively small contribution of irreducible
$2\pi$-exchange to the spin-orbit strength $F_{so}(k_f)$ for a cut-off scale of
$\Lambda = 0.65\,$GeV \cite{nucmat}. We note that without the zero-range
$\Lambda$-dependent term in eq.(33) the dashed-dotted curve would be shifted
downward by $45.7\,{\rm MeVfm}^5$ to negative values. 

\bigskip

\bild{fj.eps}{16}
{\it Fig.\,5: The strength function $F_J(k_f)$ accompanying the squared 
spin-orbit density $\vec J\,^2$ in the nuclear energy density functional 
versus the nucleon density $\rho=2k_f^3/3\pi^2$. An accurate fit of the full
line is: $F_J(k_f)= 12.80\,{\rm MeVfm}^{7/2}\cdot\rho^{-1/2} + 7.041\,{\rm  
MeVfm}^4\cdot\rho^{-1/3}$. The three horizontal dashed lines show the constant 
values $F_J(k_f)=[t_1(1-2x_1)-t_2(1+2x_2)]/32$ of the Skyrme forces MSk
\cite{pearson}, SIII \cite{sk3} and Sly \cite{sly}.}

\bigskip

In this context we mention also the relativistic $1/M^2$-correction to
$F_{so}(k_f)$ from the $1\pi$-exchange Fock diagram. Inserting the expression
$U_{ls}^{(1\pi)}(p,k_f)$ in eq.(6) of ref.\cite{uls} into the "master formula"
eq.(32) leads to the simple result: $F_{so}(k_f) = g_A^2[\ln(1+4u^2)-4u^2]/
(16Mf_\pi u)^2$, with $u=k_f/m_\pi$. As expected, this contribution with
$F_{so}(2m_\pi) = -0.86\, {\rm MeVfm}^5$ is negligibly small. The full line in
Fig.\,4 is accurately fitted by: $F_{so}(k_f)= 1.898\,{\rm
MeVfm}^3\cdot\rho^{-2/3} - 29.37\,{\rm  MeVfm}^4\cdot \rho^{-1/3}$.

Finally, we show in Fig.\,5 the strength function $F_J(k_f)$ accompanying the 
squared spin-orbit density $\vec J\,^2$ in the nuclear energy density
functional versus the nucleon density $\rho=2k_f^3/3\pi^2$.  For comparison we
have drawn the constant values $F_J(k_f)=[t_1(1-2x_1)-t_2(1+2x_2)]/32$ of the 
three Skyrme forces MSk \cite{pearson}, SIII \cite{sk3} and Sly \cite{sly} 
(dashed lines). One observes that our prediction for $F_J(k_f)$ is considerably
larger. Again, there is a strong rise of the strength function $F_J(k_f)$ as 
one goes down to very low nucleon densities $\rho < \rho_0/10$. This time the
dominant contribution comes from  the iterated $1\pi$-exchange Hartree diagram
with three medium insertions eq.(21), which gives numerically at saturation
density $F_J(2m_\pi) = 52.5\,{\rm MeVfm}^5$. It should also be noted that the 
$\vec J\,^2$-term in the energy density functional is often neglected in 
nuclear structure calculations. The full line in Fig.\,5 is accurately fitted 
by: $F_J(k_f)= 12.80\,{\rm MeVfm}^{7/2}\cdot \rho^{-1/2} + 7.041\,{\rm  MeVfm
}^4\cdot\rho^{-1/3}$, which may be useful for implementation into nuclear 
structure calculations. Note also that the $\vec J\,^2$-term in the energy 
density functional gives rise to an additional spin-orbit single-particle field
of the form $2 F_J(k_f)\vec J$. According to our calculation this additional 
spin-orbit field would be rather large and strongly density dependent.   

The full curves in Figs.\,3,5 show a strong increase as the density $\rho$
tends to zero. Although not visible, each curve approaches a finite value at 
$\rho=0$. One can analytically derive the following low density limits:
\begin{equation} \lim_{\rho \to 0} \, \rho^{-1} F_\tau(k_f) = {3g_A^2 \over
(4m_\pi f_\pi)^2} \bigg[ 1-{g_A^2  M m_\pi \over 128 \pi f_\pi^2} \bigg] =
571.3\,{\rm MeVfm}^5 \,, \end{equation} 
\begin{equation} F_\nabla(0) = {g_A^2 \over (8m_\pi f_\pi)^2} \bigg[ 3+{59 
g_A^2  M m_\pi \over 128 \pi f_\pi^2} \bigg] = 339.2\,{\rm MeVfm}^5 \,, 
\end{equation}
\begin{equation} F_J(0) = {3g_A^2 \over (4m_\pi f_\pi)^2} \bigg[ 1-{3 
g_A^2  M m_\pi \over 256 \pi f_\pi^2} \bigg] = 552.2\,{\rm MeVfm}^5 \,, 
\end{equation}
\begin{equation} F_{so}(0) = -{g_A^4  M \over \pi m_\pi (4f_\pi)^4} = 
-101.4\,{\rm MeVfm}^5 \,,  \end{equation}
to which only the diagrams with two medium insertions contribute. The large 
numbers in eqs.(34-37) arise from negative powers of the pion mass $m_\pi$ 
(so-called chiral singularities). The most singular $m_\pi^{-2}$-terms can be
traced back to the $1\pi$-exchange Fock diagram. It is important to keep in 
mind that if pionic degrees of freedom are treated explicitly in the nuclear 
matter problem the low density limit is realized only at extremely low
densities $k_f<<m_\pi/2$. Often, the opposite limit where the pion mass $m_\pi$
can be neglected against the Fermi momentum $k_f$ is already applicable at the
moderate densities relevant for conventional nuclear physics. This is
exemplified here by the approximate density dependence  $F_{\nabla,so,J}(k_f)
\sim k_f^{-1}$. Such a $\rho^{-1/3}$-behavior becomes exact in the chiral limit
$m_\pi=0$ as can be deduced by simple mass dimension counting of the dominant
iterated $1\pi$-exchange diagrams (the basic argument is that $M/f_\pi^4 k_f$
has the correct unit of MeVfm$^5$).
         
\section{Summary and outlook}
In this work we have calculated the nuclear energy density functional ${\cal 
E}[\rho,\tau,\vec J\,]$ relevant for N=Z even-even nuclei in the systematic 
framework of chiral perturbation theory. Our calculation includes the 
$1\pi$-exchange Fock diagram and the iterated $1\pi$-exchange Hartree and Fock
diagrams. These few leading order contributions in the small momentum expansion
give already a very good equation of state of isospin symmetric infinite 
nuclear matter \cite{nucmat,lutzpot}. The step to inhomogeneous many-nucleon
systems is done with the help of the density-matrix expansion of Negele and 
Vautherin \cite{negele}. Our results for the strength functions $F_\tau(k_f)$,
$F_\nabla(k_f)$, $F_{so}(k_f)$ and $F_J(k_f)$ (density dependent 
generalizations of combinations of Skyrme force parameters) are parameterfree.

We find that the effective nucleon mass $\widetilde M^*(\rho)$ deviates at most
by $\pm15\%$ from its free space value $M$, with $0.89M<\widetilde M^*(\rho)<M$
for $\rho < 0.11 \,{\rm fm}^{-3}$ and $\widetilde M^*(\rho)>M$ for higher 
densities $\rho<\rho_0 = 0.174\,{\rm fm}^{-3}$. The latter enhancement can be 
understood from Galitskii's second order calculation \cite{galitski,fetter}. 
Interestingly, a recent large scale fit of (almost two thousand) nuclide masses
by Pearson et al. \cite{pearson} finds a similarly  enhanced effective nucleon
mass: $\widetilde M^*(\rho_0) =1.05M$.

The strength of the $(\vec \nabla \rho)^2$-term, $F_\nabla(k_{f0})$, is 
comparable to that of phenomenological Skyrme forces. The magnitude of
$F_J(k_{f0})$ accompanying the squared spin-orbit density $\vec J\,^2$ comes
out larger. 

The strength of the nuclear spin-orbit interaction, $F_{so}(k_{f0})$, as given
by iterated $1\pi$-exchange is about half as large as the corresponding
empirical value $\sim 90\,$MeVfm$^5$, however, with the wrong negative sign. 
Since the (positive) contribution from irreducible $2\pi$-exchange to 
$F_{so}(k_f)$ is relatively small, there remains the challenge of understanding
the microscopic origin of the nuclear spin-orbit interaction. Evidently, chiral
perturbation theory cannot properly account for the underlying mechanisms, 
whereas relativistic scalar-vector mean-field models give a successful 
phenomenology of the nuclear spin-orbit force. Lorentz scalar and vector 
mean-fields with their in-medium behavior governed by QCD sum rules could also
provide the appropriate framework for that \cite{finelli}.

The novel density dependencies of $\widetilde M^*(\rho)$ and $F_{\nabla,so,J}
(k_f)$ as predicted by our parameterfree calculation should be explored and 
examined in future nuclear structure calculations (after adding a suitable 
positive constant to $F_{so}(k_f)$ in order to minimally repair the spin-orbit
coupling). Of course one should keep in mind that the prominent low-density 
behavior of $F_\nabla(k_f)$ as well as $F_J(k_f)$ carries little weight in the
tails of nuclear density distributions. 

For an extension to even-even nuclei with $N>Z$ the first obvious step would 
be to include the density dependent asymmetry energy $A(k_f)$
\cite{nucmat,lutzpot} (subtracted by its kinetic energy contribution) in the 
nuclear energy density functional: 
\begin{eqnarray} {\cal E}_{as}[\rho_p,\rho_n,\tau_p,\tau_n,\vec J_p,\vec J_n] 
&=&{\cal E}[\rho, \tau,\vec J\,] +{\cal E}_{coul}[\rho_p, \tau_p,\vec J_p]  
+{(\rho_n-\rho_p)^2 \over \rho} \nonumber \\ && \times \bigg\{A(k_f)-{k_f^2
\over 6M}  +{k_f^4\over 12M^3} -{5\tau k_f^2\over 56\rho M^3}\bigg\} +\dots \,,
\end{eqnarray} 
with $\rho =\rho_p+\rho_n=2k_f^3/3\pi^2$, $\tau =\tau_p+\tau_n$ and $\vec J =
\vec J_p+\vec J_n$. In an ordering scheme where one counts deviations from 
homogeneity and deviations from isospin symmetry simultaneously as small the 
energy density functional in eq.(38) would already be complete. However, such 
a formal consideration may be too simplistic in view of neutron skins, neutron
halos etc. In any case, the density-matrix expansion in eq.(1) can be
straightforwardly generalized to the isospin asymmetric situation and this way
the strength functions of terms like $[\tau_n-\tau_p+k_f^2(\rho_p-\rho_n)](
\rho_n-\rho_p)$,  $(\vec \nabla \rho_n-\vec \nabla \rho_p )^2$,  $(\vec 
\nabla \rho_n-\vec \nabla \rho_p)\cdot (\vec J_n- \vec J_p)$ and $(\vec J_n- 
\vec J_p)^2$ in the nuclear energy density functional become also accessible 
in our diagrammatic framework. For the Coulomb energy density ${\cal E}_{coul}
[\rho_p,\tau_p,\vec J_p]$ of the protons with improved treatment of the
exchange (Fock) term, see ref.\cite{titin}. 

\section*{Acknowledgement}
We thank P. Ring for suggesting this work and for useful discussions. 

\end{document}